\documentclass[11pt,a4paper]{article}

\usepackage{amsmath}%
\usepackage{amsfonts}%
\usepackage{amssymb}%
\usepackage{graphicx}
\begin{document}
\title{{\bf Generation and Amplification of Terahertz Radiation in Carbon Nanotubes}}
\author{S. S. Abukari$^a$, S.Y.Mensah$^{a,}$\thanks{Corresponding author. E-mail: profsymensah@yahoo.co.uk. Mobile: Tel.:+233 042 33837} , N. G. Mensah$^{b,\,c}$, K. W. Adu$^{d,\,e}$,\\[1ex] M. Rabiu$^f$, K. A. Dompreh$^a$, and A. Twum$^a$}
\date{}
\maketitle

\begin{flushleft}
$^a$ {\it Department of Physics, Laser and Fibre Optics Centre, University of Cape Coast, Cape Coast, Ghana}\\

\vspace{.1in}
$^b$ {\it National Centre for Mathematical Sciences, Ghana Atomic Energy commission, Kwabenya, Accra, Ghana}\\

\vspace{.1in}
$^c$ {\it Department of Mathematics, University of Cape Coast, Cape Coast, Ghana}\\

\vspace{.1in}
$^d$ {\it Department of Physics, The Pennsylvania State University Altoona College, Altoona, Pennsylvania 16601, USA}\\

\vspace{.1in}
$^e$ {\it Materials Research Institute, The Pennsylvania State University, University Park, Pennsylvania 16802, USA}\\

\vspace{.1in}
$^f$ {\it Department of Applied Physics, University for Development Studies, Navorongo Campus, Ghana.}\\

\bigskip
PACS codes: 73.63.-b; 61.48.De\\
\bigskip
Keywords: Carbon nanotubes, mathematical model, terahertz radiation
\end{flushleft}

\begin{abstract}
We investigate theoretically the feasibility of generation and amplification of terahertz radiation in aligned achiral carbon nanotubes (zigzag and armchair) in comparison with a superlattice in the presence of a constant (dc) and high-frequency (ac) electric fields. The electric current density expression is derived using the semiclassical Boltzmann transport equation with a constant relaxation time with the electric field applied along the nanotube axis. Our analysis on the current density versus electric field characteristics demonstrates negative differential conductivity at high frequency as well as photon assisted peaks. The characteristic peaks are about an order of magnitude styronger in the carbon nanotubes compared to superlattice. These strong phenomena in carbon nanotubes can be used to obtain domainless amplification of terahertz radiation in carbon nanotubes at room temperature. 
\end{abstract}

\section{Introduction}
Terahertz (THz) region of the electromagnetic (EM) spectrum refers to the frequency range between the mid infrared and the microwave region (0.3 THz - 10 THz). This EM region has a lot of promising applications in various areas of science and technologies include astronomy, broadband communication, pollution monitoring , poison gas sensing, etc. \cite{[1], [2]}. In spite of the potential applications, this region of the EM spectrum is yet to be fully exploited, due to the limited availability of device sources and detectors \cite{[3]}. Over the past decade, terahertz science and technology has advanced considerably with both optical-bench-based systems and solid state quantum cascade lasers \cite{[3]}. The main drawback of these quantum cascade THz sources  is low working temperature, making it difficult to maintain population inversion at room temperature \cite{[4]}. So far the highest temperature attained for a THz source operating at 3 THz frequency is 160 K \cite{[5]}. Creating a compact reliable source of THz radiation that could operate at room temperature still remains one of the most formidable challenges of contemporary applied physics. Most recently, carbon nanotube has been identified for THz technology due to its superb physical properties. Different proposals of carbon nanotubes for THz applications have been made; that ranges from frequency multipliers \cite{[1], [6]}, THz amplifies \cite{[7]}, switches \cite{[8]} to antennas \cite{[9]}.

Carbon nanotube (CNT)   is single layer nanometer-size cylinder made of carbon atoms. The cylinder has no bulk atoms, only surface atoms. It is a layer of graphene seamlessly rolled into cylinder. CNTs have intriguing physical properties (mechanical, electrical, optical, etc.) due to their unique structure; combine with of one-dimensional solid state characteristics with molecular dimensions \cite{[10]}. These properties vary with each CNT fundamental indices, ($n, \,m$) which specify the diameter and wrapping angle as the graphene sheet seamlessly wrapped into a CNT. As $n$ and $m$ vary, conduction ranges from metallic to semiconducting \cite{[11]}, with  $n = m$ been metallic and $m = 0$ been semiconducting. The band gap of the semiconducting CNTs  which is inverse diameter ($d$) dependent, ranges from 0.2 eV to 2 eV with $E_g = 0.9/d\, nm$ \cite{[12]}, where $d = a_o$ ($n^2 + m^2-nm$)$^{1/2}/\pi$ is the nanotube diameter and $a_o = 0.246$ nm is graphene lattice constant. The peculiarity of CNTs (e.g., strong nonparabolic dispersion law) raises the question of whether these structures could be used for high-order harmonic generation \cite{[13], [14]}. Different theoretical models and experimental techniques are being pursued to demonstrate the feasibility of CNT for THz applications\cite{[15]} - \cite{[19]}.  These ranges from THz generation Cherenkov-type emitters base on CNTs and hot electrons in quasimetallic CNT ($n-m = p$ with $p$ as a non-zero interger), frequency multiplication in chiral-nanotube-base superlattices controlled by a transverse electric field, and THz radiation detection and emission by armchair nanotubes in a strong magnetic field \cite{[15]}, ab initio molecular dynamic simulations \cite{[16]}. CNT capacitor circuit model \cite{[17]}, impedance transmission model \cite{[18]}, effective conductivity model \cite{[19]}, just to mention a few. 

Nonlinearity of the current density characteristics of CNTs gives rises to a generation of harmonics of microwaves and direct current (dc) generation as well as terahertz generation \cite{[1], [21], [22]}. However, Bloch oscillations of electrons within a CNT cause an appearance of negative differential conductivity (NDC) for dc fields larger than the critical electric field \cite{[14]}. This leads to electric current instabilities and formation of domains in the CNTs \cite{[23]}. These current instabilities are destructive to the formation of the terahertz frequency. Therefore, it is important to consider the scheme of generation of terahertz radiation than can suppress electric instabilities while preserving high-frequency gain at room temperature. Theoretical realizations of superlattice oscillators (SL) with ac bias have been reported \cite{[24]} - \cite{ [26]}, that demonstrated domainless amplification of terahertz radiation. However, there are limited reports of such effects in CNTs, especially simultaneous use of dc-ac electric field. We present here, a report on a theoretical investigation of zigzag single wall carbon nanotube (z-SWCNT) and armchair single wall carbon nanotube (a-SWCNT) in comparison with superlattice (SL) under the influence of  a time dependent dc-ac electric field for THz  frequency generation and amplification at room temperature  that indicates an order of magnitude effective suppression  of the electric instability in the one dimensinal CNTs in comparison with the two-dimensional SL  as reported in refs  \cite{[24]} - \cite{[26]}.

\section{Theoretical model}
Following the approach of refs. \cite{[15]}-\cite{[17]}, we consider an undoped a-CNT (i.e., zigzag and armchair ) exposed simultaneously to a dc-ac electric field.
\begin{equation}
 E(t)=E_o + E_1 cos\omega t \label{eq:one}
\end{equation}
where $E_o$ is the constant electric field, $E_1$ and $\omega$ are the amplitude and the frequency of the ac field respectively.The investigation is done within the semiclassical approximation in which conduction electrons with energy below the energy of the interband transitions move in the crystalline lattice like free quasi-particles with dispersion law extracted from quantum theory. Taking into account the hexagonal crystalline structure of a rolled graphene in a form of CNT and using the tight binding approximation, the energy dispersion for zigzag and armchair CNTs are expressed as in Eqns. \eqref{eq:two} and \eqref{eq:three}, respectively \cite{[1]}.
\begin{eqnarray}
	\mathcal{E}(s\Delta p_{\varphi}, p_z) &\equiv& \mathcal{E}_s (p_z ) \nonumber\\
	&=&  \gamma_0 \left[1+4cos(ap_z )cos(\frac{a}{\sqrt{3}} s\Delta p_{\varphi}) +4cos^2 (\frac{a}{\sqrt{3}} s\Delta p_{\varphi}) \right]^{1/2}\label{eq:two}
\end{eqnarray}
\begin{eqnarray}
	\mathcal{E}(s\Delta p_{\varphi}, p_z)  &\equiv& \mathcal{E}_s (p_z )\nonumber\\
	&=& \gamma_0 \left[ 1+ 4cos(ap_z )cos(\frac{a}{\sqrt{3}} s\Delta p_{\varphi}) + 4cos^2 (\frac{a}{\sqrt{3}} s\Delta p_{\varphi})\right ]^{1/2} \label{eq:three}
\end{eqnarray}
where $\gamma_0 \sim 3.0$eV is the overlapping integral, $p_z$ is the axial component of quasi-momentum, $\Delta p_{\phi}$ is transverse quasi-momentum level spacing and $s$ is an integer. The expression for a in Eqns. \eqref{eq:two} and \eqref{eq:three} is given as  $a = 3b/2\pi$, $b = 0.142$nm is the C-C bond length. The $-$ and $+$ signs correspond to the valence and conduction bands respectively. Due to the transverse quantization of the quasi-momentum, its transverse component can take n discrete values, $p_{\varphi} = s\Delta p_{\varphi} = (\pi \sqrt{3}s)/an$ ($s=1...,n$). Unlike transverse quasi-momentum $p_{\varphi}$, the axial quasi momentum $p_z$ is assumed to vary continuously within the range $0 \leq p_z \leq 2\pi/a$ , which corresponds to the model of infinitely long  of WNT ($L = \infty$). This model is applicable to the case under consideration because of the restriction to the temperatures and/or voltages well above the level spacing \cite{[1], [11]}, ie. $k_B T > \mathcal{E}_C, \Delta\mathcal{E}$, where $k_B$ is Boltzmann constant, $T$ is the temperature, $\mathcal{E}_C$ is the charging energy. The energy level spacing $\Delta\mathcal{E}$ is given by
\begin{equation}
 \Delta\mathcal{E} = \frac{\hbar v_F}{L},\label{eq:four}
\end{equation}
where $v_F$ is the Fermi velocity and L is the carbon nanotube length \cite{[5]}

Employing Boltzmann equation with a single relaxation time approximation
\begin{equation}
\frac{\partial f(p,t)}{\partial t} + eE(t)\frac{\partial f(p,t)}{\partial p}= - \frac{[ f(p,t) - f_{0}(p)]}{\tau}.\label{eq:five}
\end{equation}
where $e$ is the electron charge, $f_0 (p)$ is the equilibrium distribution function, $(p,t)$ is the distribution function, and $\tau$ is the relaxation time. The electric field $E$ is applied along CNs axis. In this problem the relaxation term $\tau$ is assumed to be constant. The relaxation term of Eqn. \eqref{eq:five} describes the effects of the dominant type of scattering (e.g. electron-phonon and electron-twistons). For the electron scattering by twistons (thermally activated twist deformations of the tube lattice), $\tau$ is proportional to $m$ (armchair dual index ($m, m$)) \cite{[26]} and the $I-V$  characteristics shows that scattering by twistons increases and decreases in the negative differential conductivity (NDC) region; the smaller $m$ is, the stronger this effect. Quantitative changes of the $I-V$ curves turn out to be insignificant in comparison with the case of $\tau = const$ \cite{[26], [27]}.
The distribution functions for zigzag CNT could be expanded in Fourier series \cite{[1]} as,
\begin{equation}
	f_0(p) = \Delta p_{\varphi}\sum_{s=1}^{n}\delta(p_{\varphi} - s\Delta p_{\varphi})\sum_{r\neq 0}f_{rs}e^{ibrp_z},\label{eq:six}
\end{equation}
\begin{equation}
	f_0(p, t) = \Delta p_{\varphi}\sum_{s=1}^{n}\delta(p_{\varphi} - s\Delta p_{\varphi})\sum_{r\neq 0}f_{rs}e^{ibrp_z}\Phi_{\upsilon}(t),\label{eq:seven}
\end{equation}
Here, the energy unit $\hbar=1$ has been used, $\delta(x)$ is the Dirac delta function, $r$ is summation over the stark component, $f_{rs}$ is the coefficient of the Fourier series and $\Phi_{\upsilon}$ is the factor by which the Fourier transform of the non-equilibrium distribution function differs from its equilibrium distribution counterpart. All other parameters are as defined above. $f_{rs}$ is expressed as.     
\begin{equation}
	f_{rs} = \frac{a}{2\pi\Delta p_{\varphi} s}\int_{0}^{\frac{2\pi}{a}}\frac{e^{-ibrp_z}}{1 + exp(\mathcal{E}_s(p_z)/k_BT)}dp_z. \label{eq:eight}
\end{equation}
Substituting  Eqns. \eqref{eq:six} and \eqref{eq:seven} into Eqn. \eqref{eq:five}, and solving with Eqn. \eqref{eq:one} we obtain
\begin{equation}
	\Phi_{\upsilon} = \sum_{k = \infty}^{-\infty}\sum_{m = \infty}^{-\infty}\frac{J_k(r\beta)J_{k- \upsilon}(r\beta)}{1 + i(\Omega r + k\omega)\tau}exp(i\upsilon\omega t),\label{eq:nine}
\end{equation}
where $\beta = eaE/\hbar$, $J_k(r\beta)$ is the Bessel function of the $k^{th}$ order and $\Omega = eaE_0$.
To obtain the current density, we expand $\mathcal{E}_s(p_z)/\gamma_0$ in Fourier series with coefficients $\mathcal{E}_{rs}$ 
\begin{equation}
	\frac{\mathcal{E}_s(p_z, s\Delta p_{\varphi})}{\gamma_0} = \mathcal{E}_s(p_z) = \sum_{r\neq 0}\mathcal{E}_{rs}e^{ibrp_z}. \label{eq:ten}
\end{equation}
Where
\begin{equation}
 \mathcal{E}_{rs} = \frac{a}{2\pi\gamma_0}\int_0^{\frac{2\pi}{a}}\mathcal{E}_s(p_z)e^{-ibrp_z} dp_z \label{eq:eleven}
\end{equation}
and expressing the velocity as 
\begin{equation}
	v_z(p_z, s\Delta, p_{\varphi}) = \frac{\partial\mathcal{E}_s(p_z)}{\partial p_z} = \gamma_0\sum_{r\neq 0}iar\mathcal{E}_{rs}e^{ibrp_z},\label{eq:twelve}
\end{equation}
the surface current density is then defined as
\begin{equation*}
j_z = \frac{2e}{(2\pi\hbar)^2}\int\int f(p) v_z(p)dp.
\end{equation*}
or
\begin{equation}
j_z = \frac{2e}{(2\pi\hbar)^2}\sum_{s = 1}^{n}\int_0^{\frac{2\pi}{a}} f(p_z, s\Delta p_{\varphi}, \Phi_{\upsilon}(t))v_z(p_z, s\Delta p_{\varphi})dp_z.\label{eq:thirteen}
\end{equation} 
The integration is taken over the first Brillouin zone. Substituting Eqs. \eqref{eq:seven}, \eqref{eq:nine} and \eqref{eq:ten} into \eqref{eq:thirteen} we determined the current density for the zigzag CNT after averaging over a period of time as
\begin{equation}
	j_z = \frac{8e\gamma_0}{\sqrt{3}\hbar na_{c-c}}\sum_{r=1}^{\infty}r\sum_{s=1}^{\infty}f_{rs}\mathcal{E}_{rs}\sum_{k=\infty}^{-\infty}\frac{j_k^2(r\beta)(1 - i(\Omega r + k\omega)\tau}{1 + ((\Omega r + k\omega)\tau)^2}\label{eq:fourteen}
\end{equation}
When the electric field amplitudes are small $\beta << 1$, we can use the Bessel function approximation $j_{\pm} \sim (r\beta/2)^2$ and $j_{0} \sim 1 - r^4\beta^2/4$. Hence from Eqn. \eqref{eq:fourteen} we obtain real part of the current density to be,
\begin{eqnarray}
	j_z &=& j_o \sum_{r=1}^{\infty}r^2\sum_{s=1}^{\infty}f_{rs}\mathcal{E}_{rs} \nonumber\\
			&&\times \left( \frac{(1+(\omega\tau)^2 - (\omega_B \tau r)^2 )}{[1 + ((\omega_B r + \omega)\tau)^2][1 + ((\omega_B r + \omega)\tau)^2]} \right)\label{eq:fithteen}
\end{eqnarray}
and the imaginary part
\begin{eqnarray}
	j_z &=& j_o \sum_{r=1}^{\infty}r^2\sum_{s=1}^{\infty}f_{rs}\mathcal{E}_{rs} \nonumber\\
			&&\times \left( \frac{i\omega\tau(1 - 3(r\Omega\tau)^2 + (\omega \tau)^2 )}{[1 + (\omega_Br\tau)^2][1 + ((\omega_B r + \omega)\tau)^2][1 + ((\omega_B r + \omega)\tau)^2]} \right)\label{eq:sixteen}
\end{eqnarray}
Where $j_o = \frac{8e\gamma_0}{\sqrt{3}\hbar an}$ and $\Omega = \omega_B$.

These are similar to the equations obtained in ref \cite{[24]} for $\beta<<1$ . On the other hand we can re-write Eqn. \eqref{eq:fourteen}  in the form of ref \cite{[25]} as;
\begin{equation}
	j_z = j_o \sum_{r=1}^{\infty}r \sum_{k=-\infty}^{\infty} J_k^2 (r\beta) j_{dc} (E_o r + kE^* ) \sum_{s=1}^nf_{rs} \mathcal{E}_{rs}\label{eq:seventeen}
\end{equation}
Where $E^* = \hbar\omega/a$ and for $k \neq 0$, $j_dc(\Omega r + kE^*)$ is the photon replicas.
Expression \eqref{eq:seventeen} is the superposition of the dc current density and its various photon replicas.  We obtained the dc differential conductivity as;
\begin{equation}
	\sigma_z = (\partial j_z)/(\partial E_o ) = j_o \sum_{r=1}^{\infty}r \sum_{k=-\infty}^{\infty} kj_k^2 (r\beta) \frac{((1 - (\Omega r + kE^* )^2 ))}{(1 + (\Omega r + kE^* )^2 )^2 } \sum_{s=1}^nf_{rs} \mathcal{E}_{rs}.\label{eq:eighteen}
\end{equation}
Where $\Omega = eaE_0$, $\beta = eaE/\omega$ for zigzag CNT and $\Omega = eaE_0/\sqrt{3}$, $\beta = eaE/\sqrt{3}\omega$ for armchair CNT.
This is exactly what is obtained by ref \cite{[26]} except that the 1D CNTs have the stark component (summation over r) which is absent in ref \cite{[26]} and consequently making the 1D CNTs more nonlinear than the 2D SL

\section{Results, Discussion and Conclusion}
We present the results of a semiclassical theory of electron transport in a CNTs simultaneously exposed to both constant (dc) and ac electric fields. The Boltzmann's equation is solved in the frame work of momentum-independent relaxation time. An analytical expression for the current density is obtained for the situation $\beta << 1$. The nonlinearity is analyzed basically on the dependence of the current density on the frequency and the d.c electric field.

\begin{figure}[thb!]
	\centering{\includegraphics[height=3.7in,width=4.9in]{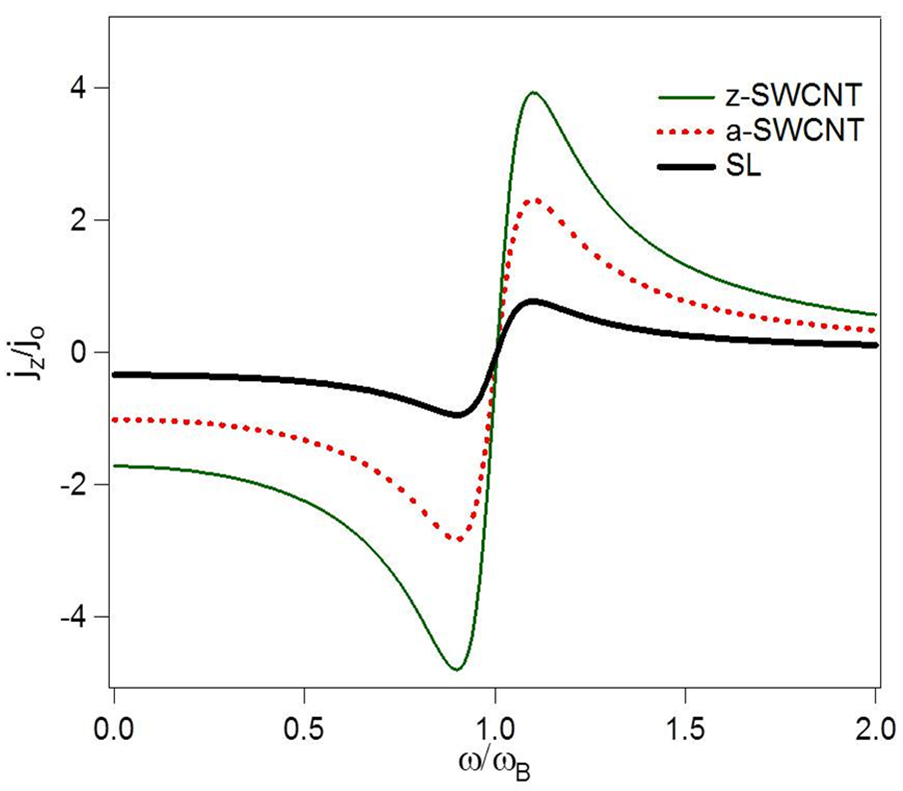}}
	\caption{$j_z/j_o - \omega/\omega_B$ curves for SL (thick solid curve) a-SWCNT (dotted curve) and z-SWCNT using $\tau =3\times10^{-12}s$, $T=287.5K$}\label{fig:DrSule1}
\end{figure}
\begin{figure}[thb!]
	\centering{\includegraphics[height=3.6in,width=4.7in]{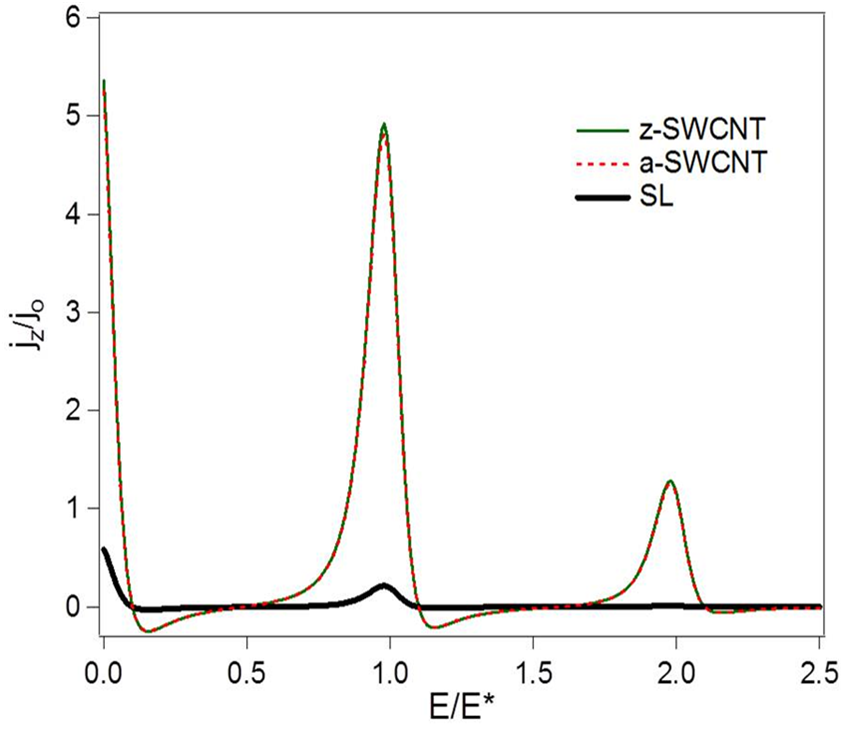}}
	\caption{$j_z/j_o - E/E^*$ curves for SL (thick solid curve) a-SWCNT (dotted curve) and z-SWCNT using   $\tau =3\times10^{-12}s$, $T=287.5K$}\label{fig:DrSule2}
\end{figure}
\begin{figure}[thb!]
	\centering{\includegraphics[height=4.0in,width=4.9in]{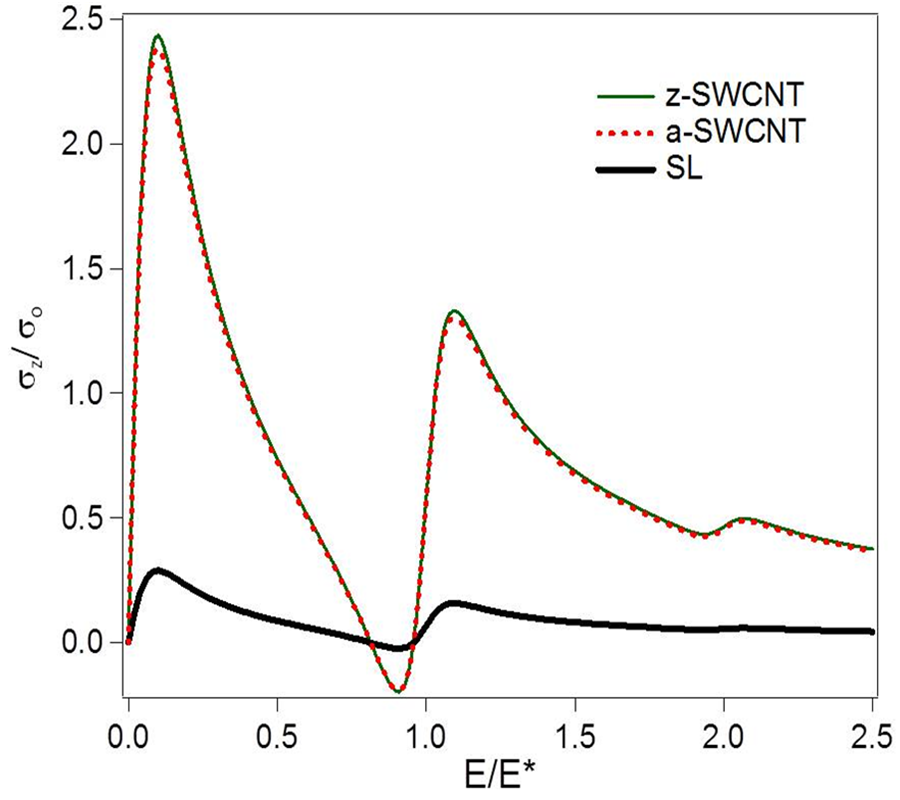}}
	\caption{$\sigma_z/j_o - E/E^*$ curves for SL (thick solid curve) a-SWCNT (dotted curve) and z-SWCNT using   $\tau =3\times10^{-12}s$, $T=287.5K$}
	\label{fig:DrSule3}
\end{figure}

In Fig. \ref{fig:DrSule1}, we show the plot of the real part of the normalized current density ($j_z/j_o$) as a function of dimensionless frequency ($\omega/\omega_B$) for a superlattice (thick solid curve), a a-SWCNT and a z-SWCNT, respectively. In all three systems, we observed that the real part of the differential conductivity is initially negative at zero frequency and becomes more negative with increasing frequency, until it researches a resonance minimum at a frequency just below the Bloch frequency ($\omega_B \tau = 10$)  and then turning positive (resonance enhancement) just below the Bloch frequency (see Fig. \ref{fig:DrSule1}). This resonance enhancement is indicative for terahertz gain at frequency without the formation of current instabilities induced by negative dc conductivity. It is worth to note that the effect is about twice and trice stronger in the a-SWCNT and z-SWCNT, respectively, in comparison to the SL, indicating stronger effective suppression of the current instability in CNTs. 

The dependence of $j_z$ on $E_o$ Eqn. \eqref{eq:seventeen} as well as $\sigma_z$ Eqn. \eqref{eq:eighteen} on $E_o$ arising under the action of strong ac field. If ac field is applied along the CNTs axis in addition to the dc field, , new transport channels are opened at large ac field which are seen as distinctive peaks in the current density and the conductivity characteristics as shown in Figs. \ref{fig:DrSule2} and \ref{fig:DrSule3}, respectively. We note a very steep positive slope in the neighborhood of the maximum electric field, which indicates that the differential conductivity is positive (PDC) in this region (see Fig. \ref{fig:DrSule2}). The PDC is considered one of the conditions for electric stability in the nonlinear system.  The negative ac conductivity at the drive frequency   appears when electric is weak. We observed that in the nonlinear regime with high-enough frequency, the dc differential conductivity is positive, while the large-signal high frequency differential conductivity remains negative (see Fig. \ref{fig:DrSule3}), which can be used for THz gain.

The magnitude of the current density as well as dc differential conductivity is an order of magnitude higher a-SWCNT and z-SWCNT in comparison to that of SL. The strong effects in CNTs are due to the higher density of free electrons in the CNTs. The mechanism of the nonlinearity in CNTs is due to the presence of the high stark components (summation with respect to $r$,  Eqns. (\ref{eq:sixteen} - \ref{eq:eighteen}) \cite{[14]} which are absent in superlattice \cite{[26]}.

In conclusion, we have used the semi-classical Boltzmann transport equation to obtain the electric current density and the dc conductivity in a achiral CNT under the influence of simultaneous dc, ac electric field applied parallel to the tube axis. Our analysis on the current density versus electric field characteristics demonstrate negative high frequency differential conductivity resonance enhancement, photon assisted peaks, suppression of current instability induced by negative dc conductivity and therefore the possibility of terahertz gain at room temperature without electric instabilities.

\end{document}